# Female leadership in software projects – A preliminary result on leadership style and project context factors

Anh Nguyen-Duc [1,2], Soudabeh Khodambashi [1], Jon Atle Gulla[1], John Krogstie[1], Pekka Abrahamsson[1, 3]

**Abstract:** Women have been shown to be effective leaders in many team-based situations. However, it is also well-recognized that women are underrepresented in engineering and technology areas, which leads to wasted efforts and a lack of diversity in professional organizations. Although studies about gender and leadership are rich, research focusing on engineering-specific activities, are scarce. To react on this gap, we explored the experience of female leaders of software development projects and possible context factors that influence leadership effectiveness. The study was conducted as a longitudinal multiple case study. Data was collected from survey, interviews, observation and project reports. In this work, we reported some preliminary findings related to leadership style, team perception on leadership and team-task context factors. We found a strong correlation between perceived team leadership and task management. We also observed a potential association between human-oriented leading approach in low customer involvement scenarios and task-oriented leading approach in high customer involvement situations.

**Keywords:** female leadership, software engineering, contingency model, project contextual factors, team coordination, team performance

## 1 Introduction

The rise of female workforce has recently gained attention as an upcoming paradigm shift in modern organizations and professional teams. Women have been found to be effective leaders in team-based, consensually-driven organizational structures that are becoming more and more popular [1, 6]. In STEM (Science, Technology, Engineering and Mathematics), however, women are still under-representative. Averaged across the globe, women accounted for less than a third (28.4%) of those employed in scientific research and development (R&D) in 2013 [4]. Although women hold close to 45% all jobs in the U.S. economy, they take less than 25% of STEM jobs [5]. The underrepresentation of women in STEM is problematic as it results in the waste of talents, gender diversity and creativity in professional workplaces.

[1] A. Nguyen-Duc, S. Khodambashi, J. Atle Gulla, J. Krogstie, P. Abrahamsson
Norwegian University of Science and Technology
Department of Computer and Information Science
e-mail:{anhn, soudabeh, jon.atle.gulla, john.krogstie, pekkaa}@ntnu.no
[2] A. Nguyen-Duc's second affiliation
University College of Southeast Norway
Department of Business Administration and Computer Science
[3] P. Abrahamsson's second affiliation
University of Jyväskylä
Faculty of Information Technology

Despite of a rich literature on gender studies, research seems still inconclusive about the reasons for the underrepresentation of women in STEM, and more importantly, how to deal with that. Many scholars believe that the existence of gender stereotypes is one of the main reasons [8-10], as the traditional male-biased practices and leadership norms function to exclude women [12]. There is an increasing attention to the gender topic in IT, i.e. Software Engineering (SE) area. A search in the Scopus database at 21st March, 2017 gave us 88 peer-reviewed publications about gender-related topics in SE area. This number is very modest compared to gender studies in other fields, i.e. Business and Management.

With the important role of software in modern informational and cyber-physical products, understanding human factors of software development, including the gender perspective, is essential to design a processes, practices and guidelines that best utilize the available workforces. To react on this above gap, we conducted a research on female leadership in software development projects, with the aim of understanding the relationship between SE-specific context factors, gender stereotype and leadership. Inheriting from a body of knowledge on gender and leadership, we do not focus on differentiating male and female leadership behaviors. Instead, we would like to know if there are project situations that best supports female leadership. Our general research question is: **How can we characterize the female leadership under different software project situations?** The paper presents our research design and preliminary results for analyzing women leadership styles and their leading situations in term of team and task characteristics.

The paper is organized as follows: Section 2 briefly presents background on leadership and gender studies, and state-of-the-art on gender study on SE. While Section 3 presents research methodology, Section 4 describes our preliminary findings and discussion. Section 5 is the conclusion and future work.

## 2   Background

### *2.1. Gender and leadership*

**Leadership:** a clear definition of what "leadership" remains somewhat elusive as numerous definitions of "leadership" has been found in literature. We found some definitions that apply well to understand female leadership in our context setting: "*behavior of an individual . . . directing the activities of a group toward a shared goal*" [43], and " *... ability of an individual to influence, motivate, and enable others to contribute toward the effectiveness and success of the organization ...* " [3]

**Gender stereotyping:** research has produced various theories about whether leadership traits and behaviors differ between men and women as distinctive biological groups. First of all, one needs to understand gender a multidimensional personality characteristic most commonly described by the difference in term of (1) biology and sex, (2) gender role, (3) causal factors and (4) attitudinal drivers [15]. Gender stereotypes are somewhat culturally shared beliefs that dictate expectations about how women and men are and how they ought to behave [20]. Thus, stereotypes can be both descriptive and prescriptive in nature. Perceived feminine competences, which are required in STEM, include, for instance, communication competences, customer and workplace relationship competences, and creativity [16, 17]. Technical competences, such as programming software architecture, are perceived as being fundamentally attached to male [18, 19].

**Leadership style**: such as relationship-oriented and task-oriented styles [30], transformational and transactional styles [1, 2, 15], and directive and participative leadership [38] have been an important topic in literature of gender and leadership. Contingency approach suggests that not only the traits and behaviors of a leader can explain for leadership effectiveness, but also the situation that the leader is at [30, 41, 42]. Fiedler developed the LPC Contingency Model, which focuses on the relationship between a trait termed the "least preferred coworker" (LPC) score and leadership effectiveness. He concluded that the most favorable situations for leaders were those in which they were well liked (good leader-member relations), directed a well-defined job (high task structure), and had a powerful position (high position power) [30, 41, 42].

**Gender and leadership:** other style category describes gender difference in term of transformational and transactional leadership [21]. Transformational leaders are characterized as inspiring, motivating, being attentive to and intellectually challenging their followers, where transactional leaders are described as contractual, corrective, and critical in their interactions with employees [21]. Female leaders were found to be more transformational than male leaders and also engaged in more of the contingent reward behaviors that are a component of transac-tional leadership [1]. Overall, researchers have asserted that there is no "*one style fits all*" solution to leadership issues and that the efficacy of various styles is contextual [22, 23]. While there are some differences between men and women when it comes to style, these differences do not lead to a clear advantage of either gender across contexts [2].

## 2.2. Gender lens in SE

SE research on gender is heavily driven by theories of gender stereotyping. In requirement engineering tasks, female performances are associated with personal factors. Less success is expected and achievements are less attributed to the own abilities [24]. Multiple case studies on women in course-based software projects characterize the collaborative learning environments for women participating [28]. The authors identified four common themes: working with others; productivity; confidence; and interest in IT careers. One key finding was that collaboration, emerging from face-to-face meetings, helps female students to build confidence via higher quality products and to reduce amount of time spent on assignments [28].

SE research also focuses on how men and women work differently regarding to specialized SE tasks. In term of coding, it is found to be different compatibility and communication levels between same gender pair and mixed gender pair [25]. Women were found to often develop deficient elements and inappropriate strategies in complex problem solving. Women tend to use bottom-up strategies while men are more risk prone and use more top-down strategies [29]. Regarding to quality assurance and testing, female developers expressed a lower level of self-efficacy than males did about their abilities to debug [26]. Further, women were less likely than males were to accept the new debugging features. Although there was no gender difference in fixing the seeded bugs, women introduced more new bugs—which remained unfixed. There was also evidence on the difference between programming environments as to which features men and women use and explore [27].

## 3. Research approach

### 3.1. Conceptual framework

Starting from the demand of understanding women's participation in SE, we gathered several factors that were found to be relevant to leadership effectiveness and team performance, as illustrated in Figure 1. The factors were found from literature as influencing factors to team performance. Based on the situational leadership approaches [30, 41], we identify factors related to leaders' characteristics:

- **Leadership style**: leadership styles, such as human-oriented style and task-oriented style [30], transformational and transactional styles [1, 2, 15], and directive and participative leadership [38] is an important variable in literature of gender and leadership. In this study, we do not focus on exploring the difference of leadership styles between women and men as it has been extensively researched. To simplify the identification of leadership style, we adopted the Fiedler LPC Contingency model [30, 41, 42].
- **Leaders' experience**: personal characteristics, such as females' interest in IT careers [28], previous experience with leadership [39], personality [39] can influence the leadership effectiveness in a given situation. It is found that if women have no task-specific experience, they do not feel confident while males apply general knowledge for specific tasks [29]. We expect to gather as much information about leadership's characteristics as possible via interviews and informal discussions.
- **Leadership self-perception** on her ability of decision making, inter-personal management and task management could give a proxy to understand leadership behavior and indirectly team performance.

Research has also shown that women are more likely to be in problematic organizational circumstances [44], as in smooth situations, agentic characteristics mattered more for leader selection, whereas in times of crisis, interpersonal attributes were deemed more important [44]. To relate the problematic situation in SE project context, we identify the product complexity, task clarity and change proneness as they have been used in literature as proxies for project context factors:

- **Product complexity**: IEEE standard defines software complexity as "*the degree to which a system or component has a design or implementation that is difficult to understand and verify*" [31]. The complexity reflects how difficult to understand the requirement, and to provide the solution.
- **Task clarity**: ambiguous requirements can lead to different interpretations among developers that lead to mistakes and confusion and consequently wasted efforts. Requirement clarity describes how clearly are requirements are presented to and understood among project stakeholders in the early phases of projects.
- **Change proneness**: characterizes how likely a requirement will change over time. For a non-experience development team, frequent changes might introduce difficulty in completing requirements and achieving customer satisfaction.

Learning from a previous work [28], we identify team dynamic as an important factor to understand female leadership behavior.

- **Team communication**: can occur in various forms, such as face-to-face meetings, pair programming and social conversations. The frequency, richness of communication is within our concern.
- **Coordination mechanism**: as a mechanism to synchronize activity and information among team members, is also related to team performance [34, 36]. We identify whether coordination mechanism used is mechanistic (task organization, task assignment, schedules, plans, project controls and specifications, routine meetings) or organic (informal discussions, developer's comments, and bug reports) [35]
- **Team relationship**: as female leadership is commonly perceived with communication competences, customer and workplace relationship competences, and creativity [16, 17], it is important to understand team members, external stakeholders and their relationships to team leaders.

**Team performance and perceived leadership effectiveness**: we identify both subjective team performance (feedbacks from team members and leader herself about how good are they as a team) and objective team performance (external evaluation of teamwork as project outcomes). Team performance evaluation is often used as a proxy of leader's effectiveness [2]. Besides, female leadership might be perceived differently due to prejudice and discrimination directed against women as leaders [6].

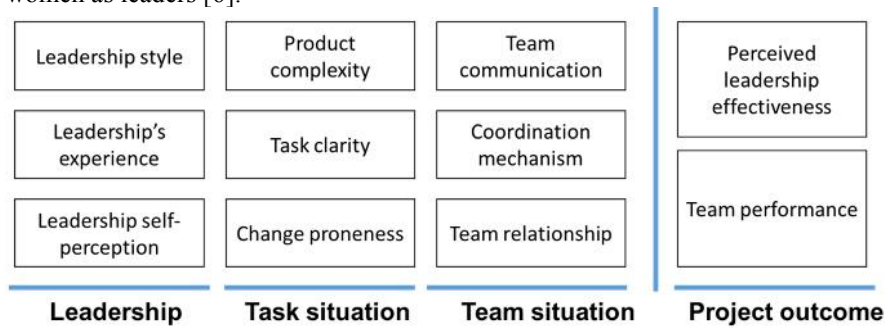

Figure 1: The conceptual framework

## 3.2. Study design

To explore the prototyping practices, we conducted a multiple exploratory case study [33]. According to Yin [33], a case study design is appropriate when (a) the focus of the study is to answer "how" and "why" questions; (b) there is probably high influence of contextual factors on the studied phenomenon. Exploratory case studies are suitable to explain the presumed causal links in real-life interventions. A multiple case study enables the researcher to explore differences within and between cases.

The underlying cases were based on software development projects in the Customer Driven Project course at the Norwegian University of Science and Technology (NTNU). Students in the course were randomly divided into groups of five to seven members. Each group has to carry out a three-month-long project for real customers from software companies, governmental agencies or research institutes. The project idea reflects a recognized need from the customer and can lead to the development of a new solution or a component of an existing one. The goal is to give the students practical experience in carrying out project development and management activities in a context as close to industry as possible. A typical project involves customers, end-users, advisors from the university and any relevant third parties. The success of the project is evaluated by a board of a course responsible, an external examiner and a team supervisor. The evaluation criteria are based on the quality of the deliverables, customer satisfaction and teamwork.

To establish the intervention, we put a female student as a leader of each project team. In Fall 2015, we observed 13 female-lead teams of 79 students. Since each student spends about 20 hours per week on the project in one semester, a typical project of 6 students corresponds to about 2,000 person hours. In the scope of this study, we will present research conducted in Fall 2015.

### 3.3. Data collection

The data collection instrument was designed to cover the conceptual elements, as described in Table 1. Data source triangulation was ensured by using both qualitative (interviews, documents, observation) and quantitative data (survey).

**Table 1: Matrix of conceptual elements and collection instrument**

| Conceptual element/ Instrument | I1 | I2 | I3 | I4 | I5 | I6 |
|---|---|---|---|---|---|---|
| Leadership style | ☐ | | ☐ | | ☐ | |
| Leaders self reflection | | | | ☐ | ☐ | |
| Leader's experience | | | | | ☐ | |
| Task clarity | | ☐ | ☐ | | | ☐ |
| Change proneness | | | ☐ | | | ☐ |
| Team collaboration | | | ☐ | ☐ | ☐ | ☐ |
| Team relationship | | ☐ | ☐ | ☐ | ☐ | ☐ |
| Team communication | | | ☐ | ☐ | ☐ | ☐ |
| Team coordination | | | ☐ | | ☐ | ☐ |
| Perceived team performance | | | | ☐ | | |

**Instrument 1**: **Leadership style survey**. We utilized the Fiedler LPC questionnaire to identify the leadership style [32]. The survey design is based on Fielder contingency model, which proposed that a leader's effectiveness is based on the match between "leadership style" and "situational favorableness" [30]. The survey responses were collected in the first two weeks of the project.

**Instrument 2**: **Project plan**. Within the course setting, each team provided their project plan with descriptions of team structure, roles, and preliminary study about the product requirements and time plan.

**Instrument 3: Team meeting observation notes**. Each team was assigned a supervisor who will assist the team from a coursework perspective. The supervisor meeting with teams offered an opportunity for research team to observe team behavior during meetings. Observation focused on team's ability to perform collective problem solving, relationship among team members and with team leader and team collaboration and coordination practices.

**Instrument 4**: **Leadership performance survey**. We designed a second survey to collect team's perception on their own teamwork and leadership. The survey used a five-point Likert scale to collect leaders and team member's opinions on (1) their own performance, (2) collective decision making, (3) team leadership and (4) task management practices.

**Instrument 5: Leader interview**. Interview with team leader was a valuable data to understand in-depth about the perception of team leader about the team performance, their experience on leading and working in SE tasks. The interview was designed as semi-structured interview that allows surprise development of the interview scenario to interesting results.

**Instrument 6**: **Final project report**. Each team delivered a 150-200 pages project report describing project planning and management, product requirement and architecture, testing and delivery. Especially, a final part of the report consisted of team reflection on project mandate, teamwork and supervision.

The process of data collection was described as in Figure 2. The data collection periods was done from Aug 2015 to Nov 2015. While I1 and I4 were mainly done by the first authors, I3 and I5 were jointly collected by all co-authors and other supervisors in the course. I2 and I6 were collected as a part of the course delivery.

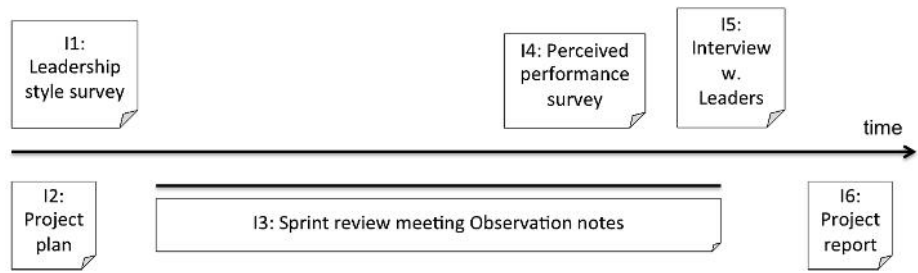

**Figure 2: Data collection process**

## *3.4. Data analysis*

At this phase, we limited ourselves in simply qualitative and quantitative analysis. The major part of analysis was done by the first author and revised by the rest.

**Narrative analysis**: is a simple form of qualitative analyzing data from i.e. interview transcripts, observation notes and textual descriptions [41]. Narratives or stories occur when one or more speakers engage in sharing an experience, which is suitable material for abstracting the project experience from our team leaders. We performed a simplified version of the analysis to extract from each project story categorical information. For instance, the level of team structure and task structure in the project will be extracted and interpreted from project report (I6). In the end of this step, we came up with a list of ordinal variables representing project context factors.

**Descriptive analysis**: quantitatively summarize characteristics of project, team and leader of our sample. The statistical summary was provided for leadership style (Section 5.2) and team reflection on leadership (Section 5.3).

**Correlation analysis**: investigates the extent to which changes in the value of an attribute are associated with changes in another attribute. Though a correlation between two variables does not necessarily result in causal effect [40], it is still an effective method to select candidate variables for causation relationship. We find Spearman's rank correlation is suitable in our case as the test does not have any assumptions about the distribution of the data and is the appropriate correlation analysis in case of ordinal variables.

## 4. Preliminary results and Discussion

## *4.1. Leadership style*

In our sample, the LPC score ranges from 47 to 109, with median value is 67, as shown in Figure 3. An interpretation approach considers the score 57 or below as a task-oriented leader and score 64 to above as a relationship-oriented leader [48]. Consequently, in general, our sample biases towards relationship-oriented leadership style. We identified five groups with task-oriented leaders, six groups with relationship-oriented leaders and two groups with mixed leaders.

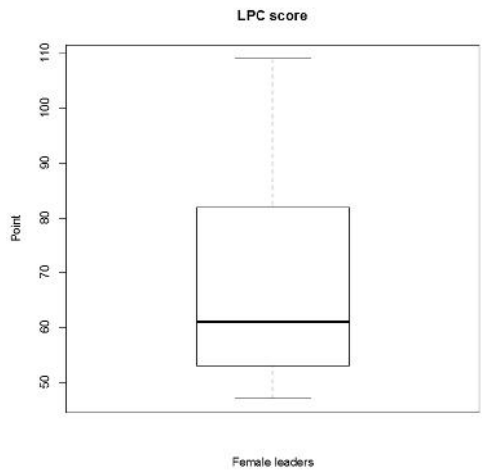

**Figure 3: Boxplot of female leaders' LPC score**

We preliminarily validate the relationship between female leadership style alone and the objective project performance:

*H0: The LPC leadership style and objective project performance are independent.*

*Test result: X-squared = 6.3254, df = 12, p-value = 0.8988*

The result of a Chi square test of independence between the two variables did not allow us to reject the null hypothesis, which infers no direct relationship between leadership style and objective team performance. This somehow aligns with our expectation, as the leadership effectiveness should be impacted by a complex combination among the leadership style and the situational favorableness.

**Table 2: Task-related contextual factors**

| Product complexity | Task ambiguity | Change proneness | No. of project |
|---|---|---|---|
| High | Medium | Low | 3 |
| High | Low | High | 1 |
| High | Low | Medium | 1 |
| Medium | High | Medium | 1 |
| Medium | Low | Medium | 1 |
| Medium | Medium | Low | 1 |
| Low | High | Medium | 1 |
| Low | Medium | Medium | 1 |
| Low | Low | High | 1 |
| Low | Low | Low | 1 |

### *4.2. Project contextual situation*

Project contextual situation is characterized by task situation and team situation. Task situation is characterized by product complexity, task clarity and change proneness, as described in Figure 1. Six projects were found to be highly complex due to the involvement of (1) multiple set of hardware devices, (2) machine learning algorithms, (3) market research and validation, and (4) industry specific technologies. 50% of the highly complex projects has medium level of task ambiguity, which introduces more challenges to the task situation. In such case, the requirement and expectation of customers were not clarified to all team members in the early phases of projects. The remaining projects (seven projects) typically involve the development of dynamic mobile/ web applications. It is noticed that the high level of ambiguity and frequent changes of requirement in later phases of projects can result in a problematic project situation regardless of product complexity.

**Table 3: Team-related contextual factors**

| Team Communication | Coordination Mechanism | Team Relationship | No of projects |
|---|---|---|---|
| High | Mechanistic | High | 1 |
| High | Mechanistic | Medium | 2 |
| High | Organic | High | 2 |
| High | Organic | Medium | 2 |
| Medium | Mechanistic | Medium | 1 |
| Medium | Organic | Medium | 3 |
| Medium | Organic | High | 1 |
| Low | Mechanistic | Medium | 1 |

Team situation is characterized by team communication, team coordination and team relationship, as described in Table 3. There are three projects found in as smooth team situations with good amount of communication among team members. Furthermore, (some of) team members knew each other before or quickly were able to set up a good information flow. There is one challenging team situation with low team communication and team members varied in term of project commitment and participation.

### *4.3. Correlation analysis*

We calculated the Spearman correlation coefficient value among variables, as shown in Figure 4. In the figure, the sizes of circles represent the coefficient value, and the color(blue or red) represents the influence direction (positive or negative). Interestingly, we found no correlation between objective team performance and perceived team performance. Leadership style is not related to team's perception on their performances or team's perception on their leadership effectiveness. Perceived team leadership is also not correlated with task.

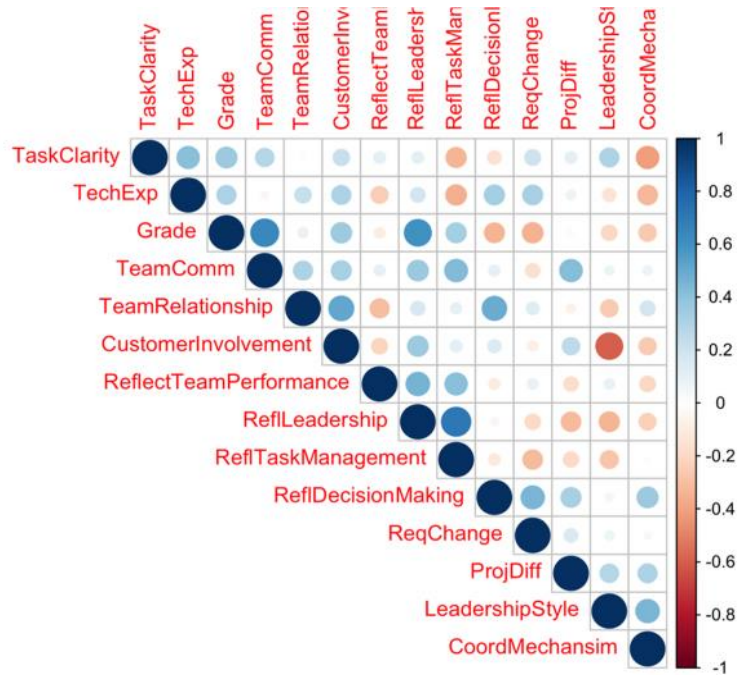

**Figure 4: Spearman correlation analysis of project context factors**

We are interested in female leadership correlations that are significant at 0.1 level, as shown in Table 4. Hopkins calls a correlation coefficient value between 0.5–0.7 large, 0.7–0.9 very large, and 0.9–1.0 almost perfect [11]. According to this scheme, objective team performance is largely correlated with team communication. The team with higher grade seems to have a better communication, in term of quality, frequency and effort spending on working together.

As shown in Table 4, team perception on team leadership is correlated with objective performance, but also with their perception on how tasks are managed. Teams with higher grades tend to be teams having positive experience with team leadership.

**Table 4: Significant correlational relationships with female leadership**

| Variable 1 | Variable 2 | Spearman Rho | P value |
|---|---|---|---|
| ObjTeamPerformance | TeamComm | 0,61 | 0,02 |
| ReflLeadership | ObjTeamPerformance | 0,50 | 0,08 |
| ReflLeadership | ReflTaskManagement | 0,53 | 0,05 |
| LeadershipStyle | CustomerInvolvement | - 0,62 | 0,02 |
| LeadershipStyle | CoordMechansim | 0,50 | 0,07 |

In the context of female leading projects, we found a strong correlation between the leadership style and customer involvement and coordination mechanisms. This of course does not exclude the cases of male leading projects. The negative coefficient value means that, in one hand, the lower score of leadership style, representing task-oriented approach is probably associated with situations of high level of customer involvement. In the other hand, the higher score of leadership style, which is human-oriented approach, can be associated with situations of low customer involvement. However, the correlation can occur because of the current customer involvement situation and assigned leadership.

## 6. Conclusions and Future work

Female participation in software industry is far less representative than that in other professional fields. This papers report a preliminary result from 13 female leading software development projects in the Norwegian University of Science and Technology. In this study, we explored leadership's characteristics and their potential relationship to project outcomes and other project context factors. Given the focus on understanding characteristics of female behavior in project leadership, the female leading projects were purposefully selected. While the comparison between male leading teams and female leading teams could give a comparable observation on leadership, it is out of the scope of this work.

Threats to validity can be discussed in term of internal, external, construct and conclusion validities [7]. Internal validity can occur when there are other influencing factors that escape our observation. We attempted to eliminate this threat by including as many factors as possible. The course coordinators (also paper authors) have run the course many years, which gave us a comprehensive insight about the courses and projects. External validity refers to the ability to generalize our project sample to industry. Projects were designed to be as close to industrial environment as possible. However, the generalization is limited by the fact that project members are students who did not spend full-time on the project. The female leadership position is also an experimented situation that might not happen naturally in industry. Construct validity relates to the transformation of conceptual elements to variables. The conceptual elements were constructed from existing literature. We also borrowed measures that were successfully adopted in literature, i.e. LPC score or team reflection survey. The transformation of textual information from data source, i.e. project reports, was cross-checked by co-authors of the paper, to reduce the bias in construct validity.

Our preliminary observation shows that the perception on team leadership is probably impacted by how the team perceive on task management. Perceived leadership and objective team performance is largely correlated, inferring one can be the indicator of the other. We also confirm that team communication is crucial for project success [34]. Last but not least, we observed a potential association between human-oriented leading approach in low customer involvement scenarios and task-oriented leading approach in high customer involvement situations. Considering the limited size of our sample, the results are not conclusive at this stage. In the future work, we will run a similar study with other customer projects to compare the results with the current study. We also plan to run different regression analysis on the whole dataset to identify indicators of female leadership effectiveness. Moreover, studying interview data with female leaders (I5) would give insight and explanation about relationships between female leadership and project factors. After having insight on female leadership in software projects, the observation will be compared and contrast with those in male leading projects.

## Acknowledgement

We would like to express our appreciation for Katja Abrahamsson, Francesco Valerio Gianni, Simone Mora, and Juhani Risku for participating in data collection.